\documentclass[11pt]{article}
  
\usepackage{epsfig,latexsym}
\newcommand{\sect}[1]{\setcounter{equation}{0}\section{#1}}

\def\aaa{angular-momentum~}

\def\al {\alpha}

\def\AAA{{\cal A}}  
 
\def\ba{\begin{eqnarray}}
 \def\bbb{background~}
\def\bbbb{backgrounds~}
\def\be{\begin{equation}}
\def\bi{\bibitem}

\def\bt {\beta}

\def\Bar {\overline}
\def\BB{{\cal B}}

\def\Bna{\Bar\na}

\def\cd{\cdot}

\def\de{\delta}

\def\ga{\dot{g}}
\def\gre{gravitational energy~}
\def\di{\partial}
\def\tLL{\dot{\LL}}
\def\tna{ \dot{\na}}
\def\tRR{\dot{\RR}}
\def\tw{\td w}
\def\tV{\dot{V}}

\def\De{\Delta}

\def\ea{\end{eqnarray}} 
\def\ee{\end{equation}}
\def\eee{equation~}
\def\eeee{equations~}
\def\em{energy-momentum~}
\def\en{energy~}
\def\ep{\epsilon}

\def \EE{{\cal E}}

\def\fr{\frac}

\def\FF{{\cal F}}

 \def\ga{\gamma}
 \def\gmn{g_{\mu\nu}}

\def\Ga{\Gamma}

\def\ha{\frac{1}{2}~}
\def\hy{hypersurface~}

\def\HH{{\cal H}}

\def\in{\infty}

\def\ka{\kappa}

\def\kkk{Killing field~}

\def\la {\lambda}
\def\lb{\label} 
\def\lll{\left(}

\def\LL{{\cal L}}
\def\LLL{\left[}

\def\me{mechanical energy~}

\def\na{\nabla}
\def\nl{\newline}
\def\nn{\nonumber}
\def\nnn{\noindent}
\def\np{\newpage}

\def\om{\omega}

\def\Om {\Omega}

\def\ra{\rightarrow}

\def\rrr{\right)}

\def\RR{{\cal R}}
\def\Ra{\Rightarrow}
\def\RRR {\right]}

\def\si{\sigma}
\def\sq{\sqrt}
\def\sqg{\sqrt{-g}}

\def\sss{spacetime~}
\def\ssss{spacetimes~}
\def\st{stationary~}

 \def\Si{\Sigma}
\def\Sc{Schwarzschild~}

\def\td{\tilde}
 
\def\tEE{\tilde\EE}
\def\tga{\tilde\ga}

\def\tf{\tilde f}

\def\ti{\times}

\def\tLL{\tilde\LL}
\def\tna{\tilde\na}
\def\tRR{\tilde\RR}
\def\tV{\td V}

\def\tw{\td w}

\def\vf{\varphi }

\def\vr{{\vec r}}

\def\vs{\vskip 0.5 cm}

\def\we{\wedge}

\def\1k{\fr{1}{\ka}}

\def\2k{\fr{1}{2\ka}}

\title{{\bf Gravitational energy in stationary spacetimes}}
\author{ Joseph Katz$^{1,2,3}$\thanks{email:
jkatz@phys.huji.ac.il} \,   Donald Lynden-Bell$^{2,3}$\thanks{email:
dlb@ast.cam.ac.uk}  \, and  Ji\v r\'\i  \,\,Bi\v c\'ak$^{2,3}$\thanks{email:
 bicak@mbox.troja.mff.cuni.cz } \\
\\ {\it $^1$The Racah Institute of Physics, Givat Ram, 91904 Jerusalem,
Israel}\\
\\{\it $^2$Institute of Astronomy, Madingley Road, Cambridge CB3 0HA,
United Kingdom}\\\\
 $^3${\it Institute of Theoretical Physics,  Charles University, 180 00 Prague 8, Czech Republic} 
\\ }
 
\begin{document}

\maketitle

\begin{abstract} 
\setlength{\baselineskip}{20pt plus2pt}
Static observers remain on Killing-vector world lines and measure the rest-mass+kinetic energies of particles moving past them, and the flux of that mechanical energy through space and time. The total mechanical energy is the total flux through a spacelike cut at one time. The difference between the total mass-energy and the total mechanical energy is the total gravitational energy, which we prove to be negative for certain classes of systems. For spherical systems, Misner, Thorne and Wheeler define the total gravitational energy this way.

To obtain the gravitational energy density analogous to that of electromagnetism we first use Einstein's equations with integrations by parts to remove second order derivatives. Next we apply a conformal transformation to reexpress the scalar   3-curvature of the 3-space. The resulting density is non-local.

We repeat the argument for mechanical energies as measured by stationary observers moving orthogonally to constant time slices like the ``zero angular momentum" observers  of Bardeen who exist even within ergospheres.
\nl
\nl
PACS numbers 04.20  -q.\,\,\, 04.20.Cv

\end{abstract} 

 \np
\setlength{\baselineskip}{20pt plus2pt}
\sect{Introduction}

In classical physics, energy has different forms which are additive. Energy is conserved and one can assess how much is transferred from one form to another. In Einstein's theory of gravitation different forms of
  energy in the matter are mixed in inseparable ways with gravitational binding energy.
Gravitational energy is diffuse, partly mixed up with various forms of matter
energy, partly stored in the gravitational field itself. Alas, the matter is the
source of gravity  so that it is impossible to disentangle gravitational energy
from other forms of energy  with or without  solving the field equations. This is a
great loss  that adds to the loss of ``gravitational  force" and   ``gravitational
energy density" with which it shares a common origin: the principle of
equivalence.

Misner, Thorne and Wheeler\footnote{Better known as MTW....} \cite{MTW} gave an expression for the total gravitational binding energy of a spherical system. They also suggest a possible definition for the \gre density in this particular case. Katz \cite{Ka2} realised that their definition of \gre could be extended to all stationary systems and gave explicit expressions for \gre density of stationary systems whose spaces were conformally flat.

Here we give an  expression for the \gre {\it density} of any \st space that is asymptotically flat. Our expression is non-local in that it involves the conformal transformation  that  removes the scalar  3-curvature of the \st 3-space   and it is invariant under transformations of spatial coordinates.    
However in specially chosen coordinates (such as isotropic ones when they exist) it may be expressed locally in terms of derivatives of the metric components. 

MTW's starting point is the difference between the \Sc mass found asympotically at large $r$ and the total mechanical energy which is the sum of the rest mass + internal + kinetic energies of the matter. This difference is the gravitational energy. Kinetic \en has to be measured relative to some standard of rest and even when that standard is agreed it will not be conserved; as fluid climbs out of gravitational wells kinetic energy decreases etc. However as fluid moves from one place to another there will be a flux of \me and in relativity we consider flux vectors describing how the \me moves from one region of \sss to another. In any \st \sss we may use the time-like Killing vector\footnote{Indices $\la,
\mu, \nu, \rho,\cdots=0,1,2,3 $; indices
$k, l, m, n\cdots=1,2,3$, the metric
$\gmn$ has signature $+---$ and $g$ is its determinant. Covariant derivatives are indicated by a $D$, partial
derivatives   by a $\di$. And for once neither $G$ nor $c$ are set equal to
$1$ and
$x^0=ct$. Finally the permutation symbol in 4 dimensions is $\ep_{\mu\nu\rho\si}$ with
$\ep_{0123}=1$ and in 3 dimensions by $\ep_{klm}$ with $\ep_{123}=1$.} $\xi^\mu$ to define `static' observers whose velocities $w^\mu=\xi^\mu/\xi$  ($\xi$ is the magnitude of $  \xi^\mu$)   are unit vectors along $\xi^\mu$.  Such observers assess the \me of a mass $m_0$ moving with 4-velocity $u^\mu$   to be  $m_0c^2/\sq{1-v^2/c^2}=m_0c^2u_\mu w^\mu$,   $v$ is its   velocity with respect to the local observer. Likewise a fluid of dust of proper rest-mass density $\rho$  (in its own rest frame) will have a rest-mass energy density in any frame moving with 4-velocity   $n^\mu$ of $\rho c^2 u_\mu n^\mu=\rho c^2/\sq{1-v_n^2/c^2}$  because the dust particles seem closer together due to Lorentz contraction of their proper volume. Such factors occur  in the following expression for rest mass.
 For a fluid of dust the rest-mass flux vector $\rho  u^\mu$ is conserved. The total rest-mass-energy is given by 
\be
M_0 c^2=\int_\Si\rho c^2 u^\mu\sqg d\Si_\mu~~~{\rm where}~~~d\Si_\mu=\fr{1}{3!}\ep_{\mu\nu\rho\si} dx^\nu \we   dx^\rho\we  dx^\si,
\lb{11}
\ee
and the 3-surface $\Si$  is {\it any} space-like \hy that spans all space. To get the \me flux vector associated with this fluid as assessed by static observers we must weight the rest-mass flux vector with the mechanical energy per unit rest-mass  $1/\sq{1-v^2/c^2}= u^\mu w_\mu$ . We obtain
\be
\rho  u^\mu u_\nu w^\nu\sqg d\Si_\mu=T^\mu_\nu w^\nu\sqg d\Si_\mu.
\lb{12}
\ee
The total \me on any cut through \sss will be
\be
E_M=  \int_\Si T^\mu_\nu w^\nu\sqg d\Si_\mu.
\lb{13}
\ee
We have motivated expression (\ref{13}) above by using a fluid made up of dust because the argument is easiest to follow in that special case;  however the final expression in term of $T^\mu_\nu$  is not at all confined to a dust fluid nor even to a fluid at all. The same expression would still hold if part of the total $T^\mu_\nu$ were the Maxwell stress \em tensor for the electromagnetic field. Likewise it would still hold for a plasma with an anisotropic pressure tensor or any other physical field.

Whereas formula (\ref{11}) holds for dust on any cut through spacetime, formula  (\ref{13}) will give answers that depend on the cut chosen unless $  T^\mu_\nu w^\nu$ is a conserved vector which is not generally so. That said, there are many important special cases for which it is conserved. Since
\be
D_\mu(  T^\mu_\nu w^\nu)=D_\mu\lll  T^\mu_\nu\fr{ \xi^\nu}{\xi}\rrr= -T^\mu_\nu w^\nu\di_\mu\log \xi,
\lb{14}
\ee
whenever   the flux vector $T^\mu_\nu w^\nu$ lies in the equipotential surfaces of constant $\xi$, it is conserved. In \st \ssss with purely toroidal motions in $(t, x^1, x^2, \vf)$ coordinates,  $T^\mu_\nu w^\nu$ is only non zero with $\mu$ in the $t$ or $\vf$ components whereas $\di_\mu\log\xi$ is only non zero with $\mu$ in the other components so in all such cases $T^\mu_\nu w^\nu$ is conserved. When that occurs, we may evaluate $E_M$ over any cut through \sss and all will give the same answer. Generally we may always evaluate $E_M$ but should not expect to get an answer independent of the cut chosen. Thus the \me seen by static observers can depend on the cut through \sss over which it is evaluated. In practice there is normally a good time coordinate such as Boyer-Lindquist in Kerr and cuts are chosen to be at constant time.

Following \cite{Ka2} the total \gre is
\be
E_G=Mc^2-E_M,
\lb{15}
\ee
where $Mc^2$ is the total energy. In this paper we convert the \gre into ``field energy" like Maxwell's $ (E^2+B^2)/8\pi$ in electromagnetism. To do this we first reexpress the $T^\mu_\nu$ in $E_M$ by using Einstein's equations. Retaining the scalar 3-curvature we remove other  second derivatives  via integrations by parts. The scalar 3-curvature is reexpressed employing a conformal transformation.  Thus the total \gre $E_G$ is found in terms of gravitational field variables. 

So far our whole argument has been developed for kinetic energy as seen by the static observers following the Killing vector field but such observers do not exist inside ergospheres. Furthermore Bardeen \cite{Ba} has emphasised that the {\it static observers} have \aaa in spaces with spin. He has introduced zero-\aaa observers (ZAMOs) who move along the normals to the hypersurfaces of constant time. Although such observers move around in azimuth they   may be preferred to   our static observers because the latter give zero motion to a static body which has \aaa (backwards).    Furthermore these   observers exist even within the ergosphere.   Thus,   $t$-hypersurface orthogonal observers move with a different 4-velocity field $\tw^\mu$ so they assess a different kinetic energy than the static observers. According to them the mechanical energy is not $E_M$ but
\be
\td E_M=\int_\Si T^\mu_\nu\tw^\nu\sqg d\Si_\mu.
\lb{16} 
\ee  
When the mechanical energy is assessed over a hypersurface of constant time, (\ref{16}) seems far more natural than our expression  (\ref{13}):   for dust the   $u^\mu\tw_\mu=1/\sq{1-\td v^2/c^2}$   factor is the same as $u^\mu n_\mu$   because the normal $n^\mu$ to the hypersurface $\Si$   is $\tw^\mu$. Writing $T^\mu_\nu =\rho  u^\mu u_\nu$ in the above expression thus yields   $(\rho /\sq{1-\td v ^2/c^2})(1/\sq{1-\td v^2/c^2})$   integrated over all space. The first '$\ga$' factor comes from the Lorentz contraction and the second from the mechanical energy per unit rest-mass. 
The factors are now the same -- they are not normally the same for static observers because for rotating systems their worldlines do not have orthogonal hypersurfaces. Thus there are strong arguments for preferring the \hy orthogonal ZAMOs to the static observers   and for considering 
\be
\td E_G=Mc^2-\td E_M
\lb{17}
\ee
 as {\it the} \gre rather than $E_G$. In axial symmetry the ZAMO mechanical energy splits into two terms `$E-\Om J$' as in classical mechanics. We discuss the \aaa density of gravitational fields elsewhere.

In what follows we derive the \gre densities for both sets of observers. These densities are different and integrate to different totals. We shall naturally   assume that the dominant energy conditions hold  \cite{HE} so that, in particular,
\be
T_{\mu\nu}w^\mu w^\nu>0~~~{\rm and}~~~T^{\mu\nu}\tw_\mu\tw_\nu>0.
\lb{18}
\ee

A proper evaluation of gravitational energy may be useful in numerical modelling of relativistic stars \cite{SG}, relativistic ellipsoidal configurations \cite{GG} and thick spherical shells \cite{CK} among others. Limits of stability are characterized by  the relative binding energy $(M-M_B)/M_B$, $M_B$ is the ``baryonic mass", which  is taken as a relativistic generalization of the $T/|W|$ ratio used in astronomy \cite{BT}. However, the more relativistic the configurations, the less representative that ratio is because $(M-M_B)c^2$ is a mixture of all forms of energy and in some models that ``binding" energy   even changes\footnote{A typical example is the binding energy of Einstein's   spherical shell of self-bound particles \cite{Ei} (see   also  \cite{CK}). A recent discussion on the sign of the binding energy in static perfect fluids can be found, for instance, in Karkowski and Malec \cite{KM}.} sign!
The $T/|W|$ ratio is better generalized as $E_K/E_G$ where $E_K$ is the  bulk kinetic energy as measured by local observers. 

 There are other   properties of \gre which will not be dealt with here but are interesting. For instance, it is not difficult to show that if \gre for static observers is extremal ($\de E_G=0$), \sss must be flat. It might be possible to show, along the lines used by Brill and Deser \cite{BD},   that $E_G$, near flatness, is a local maximum. It may be true but is as yet unclear that \gre as defined here is always negative.  We show that $E_G$ is indeed negative for static and some stationary  systems.

 \sect{Stationary spacetimes}
 
  Here is a short summary of some basic formulas for stationary \ssss in which we also fix   notations. A good recent summary of present knowledge about stationary \ssss is given in Beig and Schmidt's paper\footnote{See also \cite{ES}; the original formulation of the   projection formalism is due to Geroch \cite{Ge}.} \cite{BS}. 
\vs
\nnn{\it (i) Coordinates adapted to stationary  spacetimes, and static observers}
\vs
Stationary \ssss have a timelike \kkk $\xi^\mu$. We consider first  static observers: they have velocity components $w^\mu$ in the $\xi^\mu$ direction:
\be
w^\mu=\fr{\xi^\mu}{\xi}~~~{\rm where}~~~\xi=\sq{\gmn\xi^\mu\xi^\nu},
\lb{21}
\ee
and are at rest with respect to fixed observers at infinity.

In the projection formalism, the metric is decomposed as follows:
\be
ds^2=\lll  w_\la dx^\la  \rrr^2+\lll   \gmn-w_\mu w_\nu \rrr dx^\mu dx^\nu.
\lb{22}
\ee 
In practical calculations   the metric is most often written in  coordinates in which it takes  the following form   \cite {LL}, \cite{Li}, \cite{ES}:
\be
ds^2=\gmn dx^\mu dx^\nu=f(dx^0-\AAA_kdx^k)^2-\ga_{kl}dx^kdx^l;
\label{23}
\ee
in these coordinates,
\be
\{\xi^\mu\}=\{1,0,0,0\}~~~{\rm and}~~~\{ w^\mu\}=\{f^{-1/2},0,0,0\}.
\lb{24}
\ee       
$f,  \AAA_k$, and $\ga_{kl}$ are functions of $x^k$ only.   $\ga_{kl}$   plays a special role:  indices of tensors in $\ga$-space\footnote{Also called the ``quotient space" obtained by quotienting spacetime by the action of the stationary isometry \cite{BS}. It represents the collection of the orbits of the Killing vectors $\xi^\mu$.\nl
The $\ga_{kl}$ give a measure of proper lengths and $f$ of proper times. See a good discussion in \cite{LL}. See also \cite{Sy}.\nl Indices of $R_{\mu\nu}$ and $T_{\mu\nu}$ are always displaced with $\gmn$ not $\ga_{kl}$ or later with $\tga_{kl}$. Thus in (\ref{320}) $T^k_k=g_{k\nu}T^{k\nu}$.} will be displaced with that metric and its inverse. Thus, for instance, we shall write:
\be
\AAA^k= \ga^{kl}\AAA_l ~~~{\rm and }~~~\AAA^2 =\AAA^k\AAA_k=\ga^{kl}\AAA_k\AAA_l>0.
\label{25}
\ee
The metric components $\gmn$ and their inverse  $g^{\mu\nu}$ are:
\ba
g_{00}&=&f~~~,~~~g_{0k}=-f\AAA_k~~~,~~~g_{kl}=-\ga_{kl}+f\AAA_k\AAA_l,\label{26}\\
g^{00}&=&f^{-1}-\AAA^2~~~,~~~g^{0l}=-\AAA^l~~~,~~~g^{kl}=-\ga^{kl}.
\label{27}
\ea 
The determinants $g$ of $\gmn$ and $\ga$ of $\ga_{kl}$ are so related:
\be
\sqrt{-g}=f^{1/2}\sqrt{\ga}.   
\label{28}
\ee  
The metric   (\ref{23}) and the components $w^\mu$, (\ref{24}), keep the same form under a change  of coordinates: 
\ba
{x'}^0 &=& x^0+\tau^0(x^k)~~~ {\rm and}~~~\lb{29a}\\  {x'}^k&=&\tau^k(x^l).
\lb{29b}
\ea 
    $\AAA_k$ and $\ga_{kl}$ transform like tensors for  ${x'}^k=\tau^k(x^l)$. The interesting transformations are however those of $x^0$  that leave $f$ and $\ga_{kl}$ invariant but $\AAA_k'=\AAA_k+\di_k\tau^0$. They are associated with different $x^0=$ const hypersurfaces   which we need to fix    if we want a definite gravitational energy density. The usual gauge condition is\footnote{The gauge condition (\ref{210}) reduces the freedom of translations to such $\tau^0$ that satisfy the conditon $\na^2\tau^0=0$ [see (\ref{214}) for the definition of $\na^2$]. If we ask for $\tau^0=$ const at infinity this fixes $x^0$ up to a constant.}
\be
\di_k(\sq{\ga}\AAA^k)=0.
\lb{210}
\ee
The dominant terms of the asymptotic solutions for isolated sources are given in MTW  \cite{MTW}  on page 456. In properly chosen asymptotically  Minkowski orthogonal coordinates at infinity, 
 \be
f\ra1 - \fr{2m}{r}~~~,~~~ \AAA_l\ra  - 2\fr{\ep_{lmn} n^m j^n}{r^2}~~~,~~~\ga_{kl}\ra  \de_{kl}(1+ \fr{2m}{r})~~~{\rm where}~~~r= \sqrt{x^2+y^2+z^2}\ra \in,
\label{211}
\ee
and
\be
 n^k=\fr{x^k}{r}~~~,~~~\sum_k (n^k)^2=1~~~,~~~m=\fr{GM}{c^2}~~~,~~~j^k=\fr{GJ^k}{c^3}.
 \label{212}
\ee  
$Mc^2$ is the total energy of spacetime and $\vec J=\{ J_k\} = \{ - J^k\} $     its angular momentum vector.
 
  Einstein's equations,
 \be 
 R_{\mu\nu}=\ka(T_{\mu\nu}-\ha\gmn T )~~~{\rm with}~~~\ka=\fr{8\pi G}{c^4},
 \label{213}
 \ee
  are better written in terms of projected components that are gauge invariant:
\be
R_{\mu\nu}w^\mu w^\nu~~,~~R_{\mu\nu}(g^{\mu\rho}-w^\mu w^\rho)(g^{\nu\si}-w^\nu w^\si)~~{\rm and} ~~~R_{\mu\nu}(g^{\mu\rho}-w^\mu w^\rho)w^\nu.
\lb{213b}
\ee
  In these expressions in our coordinates,  $\AAA_k$ must necessarily appear  in a   curl. 
   Let  $\na_k$ represent covariant derivatives in $\ga$-space; we shall write 
  \be
  \ga^{kl}\na_l=\na^k~~~{\rm and}~~~\na^k\na_k=\na^2.
  \label{214}
 \ee
 The gauge invariant components of the Ricci tensor,   invariant for time translations (\ref{29a}) and covariant for spatial coordinate transformations (\ref{29b}),  are equivalent to \cite{LL}\footnote{In the 1959 edition of Landau and Lifshitz there is apparently a printing mistake in the third equation on page 301. What in their notations is written $+\fr{h}{2}f^{\al\ga}f_{\bt\ga}$ should be $-\fr{h}{2}f^{\al\ga}f_{\bt\ga}$.}: 
 \ba
R_{00}&=& {\ha}\na^2f-\fr{1}{4f}\na^kf\na_kf+  f^2 \na^{[k}\AAA^{l]} \na_{[k}\AAA_{l]},
\label{215}\\
R^{kl}=g^{k\mu}g^{l\nu}R_{\mu\nu}&=& -{\fr{1}{2f}}  \na^k\na^l f+\fr{1}{4f^2}\na^kf\na^lf+2f\ga_{mn} \na^{[k}\AAA^{m]} \na^{[l}\AAA^{n]}+\RR^{kl},
\label{216}\\
R_0^k=g^{k\nu}R_{0\nu}&=&- \fr{3}{2}\na_lf  \na^{[k}\AAA^{l]}-f\na_l(\na^{[k}\AAA^{l]}),    
\label{217}
\ea
  and the 4-scalar curvature
\be
R=f^{-1}R_{00}-\ga_{kl}R^{kl}.
\lb{217b}
\ee 
In (\ref{216}),  $\RR^{kl}$ are the components of the Ricci tensor of the $\ga$-space,  {\it not} of the $x^0=$ const hypersurface.
The following  linear combination of  (\ref{215}) and  (\ref{216}) is of particular interest:
\be
-\ga_{kl}R^{kl}-f^{-1}R_{00}=-\ga_{kl} \RR^{kl} -3f \na^{[k}\AAA^{l]} \na_{[k}\AAA_{l]}=-\RR  -3f \na^{[k}\AAA^{l]} \na_{[k}\AAA_{l]};
\label{218}
\ee
$ \RR$ is the scalar curvature of  the   $\ga$-space.
The left hand side of  (\ref{218}) can   be rewritten  using Einstein's \eeee (\ref{213}):
\be
-\ga_{kl}R^{kl}-f^{-1}R_{00}=-2\ka T^0_0+2\AAA_k R^k_0.
\label{219}
\ee
Therefore,   (\ref{218}) and (\ref{219}) imply 
\be
-\RR -3f \na^{[k}\AAA^{l]}  \na_{[k}\AAA_{l]}-2\AAA_k R^k_0 =-2\ka T^0_0.
 \label{220}
\ee
The last two terms on the left hand side can be recast in another form using (\ref{217}):
\be
-3f \na^{[k}\AAA^{l]}  \na_{[k}\AAA_{l]}-2\AAA_k R^k_0=-\na_{[k}(f \AAA_{l]}) \na^{[k}\AAA^{l]} - \na_k\lll 2f\AAA_l \na^{[k}\AAA^{l]}\rrr.
\lb{221}
\ee
Inserting this new expression   into (\ref{220}),  then   transferring $ -\na_k\lll 2f\AAA_l \na^{[k}\AAA^{l]}\rrr$ from left to right we get
\be
-\RR -\na_{[k}(f \AAA_{l]}) \na^{[k}\AAA^{l]}   
= -2\ka T^0_0+ \na_k\lll 2f\AAA_l \na^{[k}\AAA^{l]}\rrr.
\lb{222}
\ee 
Before going further, it is perhaps interesting to rewrite Einstein's equations  in terms of ``gravoelectric" and``gravomagnetic" vector fields $\EE_k$ and  $\BB^m$   living in   $\ga$-space in the spirit of \cite{LL}, \cite{LN}, \cite{Na} and \cite{TPM}. The gravoelectric potential
\be 
\psi = \log\sq{f},
\label{223}
\ee
and the gravomagnetic vector potential is $\AAA_k$.
   $\EE_k$  is defined by:
\be
\EE_k= - \di_k\psi ~~~{\rm and}~~~\ga^{kl}\EE_k\EE_l=\EE^2>0.
\label{224}
\ee
 $\BB^m$  is divergenceless\footnote{In \cite{LL} and \cite{Na}, $\BB$'s are replaced  by  $\HH=f^{1/2}\BB$. This $\HH$ is called $\BB$ in \cite{Ka2}.}:
  \be
   (\di_{k}\AAA_{l} - \di_{l}\AAA_{k} )=\eta_{klm}\BB^m~~~{\rm or}~~~\BB^m=  \eta^{mkl} \di_{k}\AAA_{l}~~~{\rm with}~~~\eta_{klm}=\sqrt{\ga}\ep_{klm}; 
\label{225}
 \ee
in particular,
 \be
   \na_{[k}\AAA_{l]} \na^{[k}\AAA^{l]}= \fr{1}{4}\eta_{klm}\eta^{kln}\BB^m\BB_n=\ha\BB^m\BB_m=\ha\BB^2>0.
\lb{226}
 \ee
  In terms of $\EE$'s and $\BB$'s the  components of the Ricci tensor take the following forms:
 \ba
 f^{-1}R_{00}&=&-\na\cdot\EE+\EE^2+\ha f\BB^2\label{227},\\
 R^{kl}&=&+\na^k\EE^l-\EE^k\EE^l+\ha f\lll   \ga^{kl}\BB^2-\BB^k\BB^l    \rrr+\RR^{kl},
\label{228}\\
 f^{-1}R^k_0&=&\ha \eta^{klm}\lll   -\na_l\BB_m+3\EE_l\BB_m    \rrr=\ha \{ -\na\ti\BB+3~\EE\ti\BB \}^k.
 \label{229}
  \ea 
  Also,
  \be
  R=-2\na\cdot\EE+2\EE^2-\ha f\BB^2-\RR,
  \ee
and   equality (\ref{222}) takes this form:
\be
-\RR -  f\lll\ha\BB^2+ \BB\cdot\AAA\ti\EE \rrr    
= -2\ka T^0_0+ \na\cdot  \lll f\AAA \ti\BB\rrr.
\lb{230}
\ee  
Dots and cross products are defined in $\ga$-space. 
\vs
\nnn{\it (ii) Metric components adapted to   observers with velocities orthogonal to constant time slices}
\vs
The coordinates are the same as in {\it (i)} but observers are different. 
They   have different 4-velocity components;   we denote them by  $\tw^\mu$ instead of $w^\mu$.   $\tw^\mu$  is associated with the $1+3$ decomposition  familiar in the Hamiltonian formulation of Einstein's equations\footnote{The original reference is Dirac \cite{Di}. See also \cite{ADM}. Modern versions are given in various books like \cite{MTW} and \cite{Wa}. The components $W^k$ in (\ref{233}) correspond to the shift vector $N^k$ and $\td f^{1/2}$ to the lapse $N$.}.  By definition, coordinates exist in which 
\be
\{ \tw_\mu\}=\{\td f^{1/2},0,0,0\}.
\lb{231}
\ee  
The metric (\ref{22}) with $\tw_\mu$ replacing $w_\mu$   now takes the following form:
\be
ds^2=\tf (dx^0)^2-\tga_{kl}\lll   dx^k-W^kdx^0  \rrr  \lll   dx^l-W^l dx^0  \rrr,  
\lb{232}
\ee 
in which $W^k$ are the components of the local coordinate  velocities in units of $c$:
\be
W^k= \fr{\tw^k}{\tw^0}=\fr{dx^k}{dx^0}; ~~~{\rm also~ set}~~~W^2=\tga_{kl}W^kW^l. 
\lb{233}
\ee
The metric components and their inverses are:
\ba
g_{00}&=&\tf-W^2~~~,~~~g_{0l}=W_l=\tga_{lk}W^k~~~,~~~g_{kl}=-\tga_{kl},\lb{234}\\
g^{00}&=&\tf^{-1}~~~,~~~g^{0l}=\tf^{-1}W^l~~~,~~~g^{kl}=-\tga^{kl}+\tf^{-1}W^kW^l.
\lb{235}
\ea
The determinant $g$ is related to the determinant  $\tga$ of $\tga_{kl}$ like this:
\be
\sq{-g}=\tf^{1/2}\sq{\tga}.
\lb{236}
\ee
The   metric (\ref{232}) and the velocity components  (\ref{231}) keep the same form under a change of coordinates (\ref{29a}) and (\ref{29b}). Some of
   Einstein's equations are now better written in terms of  Einstein's tensor components $G^{\mu\nu}$:
\be
G^{\mu\nu}= R^{\mu\nu}-\ha g^{\mu\nu} R=\ka T^{\mu\nu}.
\lb{237}
\ee
The simplest forms belong to the projected components of $G^{\mu\nu}$, see (\ref{213b}), with $w^\mu$ replaced by   $\tw^\mu$. In our coordinates,   \cite{ADM}\footnote{The reference contains   the clearest and most detailed calculations. We rewrote the formulas in our notations.}: 
\ba
2\tf G^{00}&=&   \tRR+K^2-K^{kl}K_{kl},    \lb{238}\\
 R_{kl}~&=&  \tRR_{kl}+ \tna_{(k}\tEE_{l)}-\tEE_k\tEE_l   + 2\tf^{1/2}K^m_{(k}\tna_{[l)} W_{m]} -\tf^{-1/2}\tEE_mW^mK_{kl}        \nn \\
 &&~~~~~~~~~~~~~~~~~~~~~~~~~~~~~~~~~~~~~~~~~~~~~~+\tna_m\lll   \tf^{-1/2}W^m  K_{kl}  \rrr, \lb{238b}\\ 
\tf^{1/2} G^0_k&=&\tf^{1/2} R^0_k=  -\tna_l \lll   K^l_k-\de^l_k K   \rrr,  
\lb{238c}
\ea
and
\be
R=2(\tf G^{00}-\tga^{kl}R_{kl}).
\ee
In these formulas $\tRR_{kl}$ is the Ricci tensor of the spatial metric $\tga_{kl}$; $\tRR=\tga^{kl}\tRR_{kl}$ the corresponding scalar curvature while  $K_{kl}$   (the second fundamental form of $x^0=$ const) is:
\be
K_{kl}=\tf^{-1/2}\tna_{(k}W_{l)}~~~ ; ~~~ K = K^k_k=\tga^{kl}K_{kl}~~~{\rm while}~~~\tEE_k=-\di_k\log\tf^{1/2}. 
\lb{239}
\ee
 $\tna_k$ is a $\tga$-covariant derivative. The time coordinate must be   fixed if we wish to integrate over a definite spacelike hypersurface. It is fixed up to a constant by a gauge condition like (\ref{210}). The most common choice in the Hamiltonian formulation and in numerical relativity are maximal spacelike hypersurfaces $K=0$ which imply that observers are expansion free, $D_\mu\tw^\mu=0$. In stationary \ssss the condition reduces to 
\be
 \tna_kW^k=0  ~~~{\rm  equivalent~to}~~~K=0,
  \lb{240}
\ee 
as follows form (\ref{239}). Notice that the condition is identically satisfied   in axial symmetry [see (\ref{413}) below].
\vs
{\it(iii) Relations} 
\vs
The metrics in the form (\ref{23})-(\ref{27}) and (\ref{232})-(\ref{235}) are the same metrics in the same coordinates, adapted to different decompositions associated with different families of observers. The $\tf, W^k$ and $\tga_{kl}$ are related as follows to the $f, \AAA_k$ and $\ga_{kl}$:
\be \tf=\fr{f}{1-f\AAA^2}~~,~~ W^k=-\fr{f\AAA^k}{ 1-f\AAA^2}~~,~~\tga_{kl}=\ga_{kl}-f\AAA_k\AAA_l,
\lb{241}
\ee
and reciprocally,
\be
f=\tf-W^2~~~,~~~\AAA^k=- \tf^{-1}W^k~~~,~~~\ga_{kl}=\tga_{kl}+f\AAA_k\AAA_l.
\ee
Notice that $\AAA^k=\ga^{kl}\AAA_l$,  not $\tga^{kl}\AAA_l$.
We now turn to the calculation of gravitational energy.
\sect{Gravitational energy}
\nnn{\it (i) As calculated by static observers}
\vs
If $w^\la$ are the velocity components of static observers, the mechanical energy   they evaluate on $x^0$ = const is, as pointed out in (\ref{13}),
\be
  E_M=\int_{x^0} T^\mu_\nu w^\nu \sq{-g}d\Si_\mu=\int_{x^0} T^0_0 dV~~~{\rm where} ~~~ dV=\sq{\ga}d^3x=(1-f\AAA^2)^{-1/2}d\tV.
\lb{31}
\ee
The right hand side is written in our special coordinates. $dV$ is the proper volume element in $\ga$-space  and not of the $x^0=$ const hypersurface which is $d\tV$. These two coincide in static spacetimes.  

Now consider (\ref{230}). Divide by $2\ka$, multiply by $\sq\ga$ and integrate over the whole space. The boundary conditions imply that the integral\footnote{We use abundantly Stokes theorem  \cite{Wa}, or Gauss theorem \cite{LL}, or Gauss-Ostrogradsky theorem \cite{GB} or Green-Ostrogradsky theorem \cite{Gi} or Green's theorem \cite{La} or the divergence theorem \cite{BT}. Hadamard \cite{Ha} gives it no name and regards it as a generalization of  $\int_a^bf'(x)dx=f(b)-f(a)$. It is   useful to remember  that the divergence theorem applies to divergences of continuous and single valued functions.} of $\2k \di_k\lll \sq{\ga} f\AAA\ti\BB\rrr$ is zero. Notice that if the gauge condition  (\ref{210}) was not fixed,  $\2k \di_k\lll \sq{\ga} f\AAA\ti\BB\rrr$ would not be fixed and the remaining integrand in  (\ref{32}) would also be gauge dependent.   It follows now  that the gravitational energy, see (\ref{15}),
\be
E_G=Mc^2-E_M=Mc^2-\2k\int_{x^0} [\RR+  f(\ha\BB^2+ \BB\cdot\AAA\ti\EE)] dV.
\lb{32}  
\ee   
We shall next eliminate the $Mc^2$ from (\ref{32})   and obtain an integral that depends entirely on field components and their derivatives.\footnote{Notice that on a general \hy in a  general \sss one may specify the lapse function $f^{1/2}$ arbitrarily by arbitrary time reparametrizations ${x'}^0=\tau^0(x^0)$. However, in stationary \ssss there are privileged coordinates in which the metric has the form (\ref{23}) and the allowed transformations of coordinates are those given by (\ref{29a}) and (\ref{29b}). Our gauge condition (\ref{210}) fixes the spacelike hypersurface.}  To this effect,
 we make a conformal transformation of the $\ga$-metric, say,
\be
\ga^*_{kl}= {\rm e}^{2\chi}\ga_{kl}, 
\label{33}
\ee
and set, by analogy with (\ref{224}),
\be
\FF_k=-\di_k\chi;~~~{\rm also}~~~\ga^{kl}\FF_k\FF_l=\FF^l\FF_l= \FF^2.
\label{34}
\ee
The conformal scalar curvature $\RR^*$ and $\RR$ are so related:
\be
  {\rm e}^{2\chi}\RR^*=\RR+4\na_k{\FF}^k-2{\FF}^2.
\label{35} 
\ee
 We then define  $\chi$  by the condition that $\RR^*=0$, that is
\be
 \RR= -4\na_k{\FF}^k+2{\FF}^2,    
\label{36}
\ee
or
\be
\na^2{\rm e}^{\chi/2}-\fr{1}{8}\RR{\rm e}^{\chi/2}=0.
\lb{37}
\ee
The $\ga$-metric becomes spherical at large distance and all spherical metrics are conformally flat. Therefore we shall ask the factor   ${\rm e}^{-2\chi}$ to behave   like the conformal factor of  $\ga_{kl}$ near infinity  as in (\ref{211}): 
\be
{\rm e}^{-2\chi}\ra 1+\fr{2m}{r}~~~{\rm or}~~~ \chi\ra-\fr{m}{r}~~~{\rm for}~~~r\ra\in. 
 \lb{38}
\ee
  Equation (\ref{37})  has been considered by Cantor and Brill \cite{CB} who found a necessary and sufficient condition for an asymptotically flat metric to be conformally equivalent to another asymptotically flat metric with zero scalar curvature. They showed that the condition for the existence and uniqueness of a solution of (\ref{37}) is that 
\be
\int\na q\cdot\na q \,dV\ge -\fr{1}{8}\int q^2\RR dV
\lb{39}
\ee 
for all smooth functions $q\ne 0$ of compact support.
Notice that (\ref{220}) with $R^k_0=\ka T^k_0$ can also be written as
\be
\RR+\fr{3}{2}f\BB^2=2\ka f^{-1}T_{00}.
\lb{310}
\ee 
This shows  that in non-static spacetimes outside matter $\RR=-\fr{3}{2}f\BB^2<0$ and conditions (\ref{39}) are not automatically satisfied.\footnote{The original paper \cite{CB} contains    precise statements about this theorem and useful references. It gives an example   of an axially symmetric space in which the solution of (\ref{37}) does not exist. The reason it does not exist is that $\RR$ is not positive everywhere.}

If a solution of (\ref{37}) exists, we may  replace $\RR$ in (\ref{32}) by its equivalent  (\ref{36}) and notice that because of (\ref{38}) 
\be
\fr{1}{2\ka}\int \na_l(4\FF^l) dV=  \fr{2}{\ka}\int_{r\ra\in}\FF^l dS_l= \fr{2}{\ka}\lll   - \fr{m}{r^2} \rrr 4\pi r^2=- Mc^2.
\label{311}
\ee 
As a consequence of  (\ref{311}), $Mc^2$ disappears from (\ref{32}) and we obtain for $E_G$:
\be
E_G =-\1k\int_{x^0} \{\FF^2+  \fr{1}{4}f [ \BB^2 +2 \BB\cdot(\AAA\ti\EE) ] \} dV=   \int_{x^0}\ep_G d\tV;
\lb{312}  
\ee
we used (\ref{31}) to reexpress $dV$ in terms of the proper volume element $d\tV$ on the slice $x^0$ = const.
With the gauge condition (\ref{210}), the \gre {\it density} $\ep_G$  on the fixed $x^0=$ const hypersurfaces   is well defined. It is a scalar for spatial coordinate transformations. Notice that it is not purely local as it involves   a quantity, $\FF^2$, defined by the solution of an elliptic \eee (\ref{37}). 
 
 In practical calculations of the total gravitational energy,   (\ref{312}) is not so good because one needs to solve \eee (\ref{37}). We  now give an expression for $E_G$   in terms of metric components and their derivatives. That expression does not depend on the existence of a solution of equation (\ref{36}). 
To this effect, we first notice that calculations near infinity are often simpler in spherical coordinates rather than $x,y,z$ or axial coordinates $R, z, \vf$ like in section 4 {\it(iii)}. If the metric of the limiting flat space   has   a general form  
\be
d\Bar s^2=({dx^0})^2-\Bar \ga_{kl}dx^kdx^l,
\label{313}
\ee
it is useful to take $d\Bar s^2$ as a flat ``background metric".\footnote{The method that consists in introducing a second metric avoids the non-covariance and allows the use  of coordinates that are not Minkowski coordinates at spatial infinity. The method goes back to a paper of Rosen \cite{Ro}. Details on how   suitable background metrics and mappings between the \bbbb and physical \ssss may be introduced, in particular in asymptotically flat spacetimes, can be 
 found in   \cite{BK}; see also \cite{KBL}.} This is done by   replacing in $\RR_{kl}=\ga_{km}\ga_{ln}\RR^{mn}$ partial derivatives $\di_k$ by $\Bna_k$-covariant  derivatives in the $\Bar\ga$-background    and  the $\ga$-Christoffel symbols $\Ga^m_{kl}(\ga)$ by the tensors 
\be
\De^m_{kl}=\Ga^m_{kl}(\ga)-\Bar\Ga^m_{kl}=\ha \ga^{mn}\lll \Bna_k\ga_{nl}+\Bna_l\ga_{nk}-\Bna_n\ga_{kl}      \rrr.
\lb{314}
\ee
 $\Bar\Ga$'s are Christoffel symbols  in   $\Bar \ga$-space. Thus we write $\RR_{kl} $ as follows:
\be
\RR_{kl}=\Bar\na_m\De^m_{kl}-\Bar\na_k\De^m_{lm}+\De^m_{kl}\De^n_{mn}-\De^m_{kn}\De^n_{ml}.
\lb{315}
\ee 
 Now remember  a   familiar  identity usually used  to construct   Einstein's  Lagrangian:\footnote{Notice the first $\na_k$ and the second $\Bna_k$.} 
\be
   \RR =\LL   - \na_lk^l~{\rm where}~ k^l =   \ga^{-1}\Bna_k(\ga \ga^{kl})~~~{\rm and}~~~\LL=-\ga^{kl}\lll   \De^m_{kl}  \De^n_{mn}-  \De^m_{kn}  \De^n_{ml}     \rrr.
\label{316}
\ee
 We then substitute $  \RR$ given by (\ref{316})  into  (\ref{32}). The asymptotic conditions on the metric have the result that, like in (\ref{311}),
\be
 \2k\int \na_lk^ldV=-Mc^2.
\lb{317}
\ee
Thus (\ref{32}) may now be written as 
\be
E_G= -\2k\int_{x^0} \{ \LL+    f[ \ha\BB^2 +   \BB\cdot(\AAA\ti\EE) ]\} dV.
\lb{318}  
\ee
$dV\!$,  as should be remembered, is not the proper volume element on $x^0=$ const [see (\ref{31})].      
$\LL$ is   a scalar  density because the $\De$'s are tensors. However, their value depends on the mapping on the flat \bbb which is equivalent to choosing special coordinates.   $E_G$ can be calculated with this expression in terms of  the metric components and their first order derivatives and is    correctly given by the integral. Notice  that if, for instance,  $E_G$ is found to be negative, then gravitational energy is negative irrespective of the choice of coordinates and irrespective of the nature of the sources of gravity. The formula is valuable in this respect.   (\ref{318}) is another form of the general formula obtained in \cite{Ka2} (see the Appendix).   

  It is at first surprising that the integrand does not contain derivatives of $f$ because in the Newtonian approximation it is proportional to the square of $\di_k f$.  But linearization is usually done in harmonic coordinates which, in particular, imposes (with Minkowski coordinates in the background):
\be
\di_l(\sqrt{-g}g^{kl})=0, ~~~ {\rm or~  equivalently}~~~ \fr{1}{\sq\ga}\di_l(\sq{\ga}\ga^{kl})= -\fr{1}{2f} \ga^{kl}\di_l f=\EE^k.
\label{319}
\ee 
This is how  $\di_k f$   reappears in $E_G$ in the Newtonian limit.  
 
It is also possible to write the total gravitational energy in terms of field components and a {\it covariant} integrand with the non-locality ``shuffled into" a matter tensor integral. To that effect consider (\ref{227}) and (\ref{228}) in the following combination:
\be
\ga_{kl}R^{kl}-3f^{-1}R_{00}=2\ka T^k_k= \RR+4\na_l\EE^l-4\EE^2 -\ha f\BB^2. 
\lb{320}
\ee
Replace $\RR$ by $\LL-\na_lk^l$, while moving the $\EE$ and $\BB$ terms to the left hand side:
\be
4\EE^2 +\ha f\BB^2+2\ka T^k_k=\LL+\na_l (4\EE^l-k^l).
\lb{321}
\ee
The volume integral of the divergence on the right hand side is equal to zero because of the asymptotic conditions as follows from (\ref{311}) and (\ref{317}).
Thus, after dividing by $2\ka$, the integral of $\LL$ reduces to
\be
\2k\int\LL dV=\1k\int (2\EE^2+\fr{1}{4}f\BB^2) dV + \int T^k_k dV.
\lb{322}
\ee
Replacing the $\LL$ integral in (\ref{318}), we obtain  a formula for $E_G$ in which the non-local character appears in the matter tensor part $T^k_k$:
\be
E_G=-\1k\int_{x^0} [2\EE^2+ \ha f(\BB^2+ \BB\cdot\AAA\ti\EE )] dV-\int_{x^0} T^k_k dV.
\lb{323}
\ee  
In particular, for   perfect {\it static} fluids ($\AAA=\BB=0$)  with local pressure $p$ ($T^k_k=-3p$):
 \be
E_G=-\fr{2}{\ka}\int \EE^2dV +3\int pdV. 
\label{324} 
 \ee
In the weak field limit, the first term represents twice the gravitational energy while, thanks to the virial theorem, the second term is minus once the same. 

The important formulas of this subsection are: (1) the gravitational energy {\it density} as calculated by observers at rest    given in (\ref{312}) in terms of a {\it non-local} field (which may not always exist), (2) the {\it total} gravitational energy expressed in  terms of {\it local fields} given in (\ref{318}), (3) the total  \gre calculated in terms of {\it local fields} with a real density,  the non-locality  being  given  in  terms of  the source of gravity, (\ref{323}).

\vs
\nnn{\it (ii) Gravitational energy   for   observers moving orthogonal to constant time slices}
\vs
The mechanical energy is now denoted $\td E_M$ because it is   different from $E_M$. With  $\tw_\mu$, see (\ref{231}), and the metric, see (\ref{232}),  
\be
\td E_M=\int_{x^0}  T^\mu_\nu \tw^\nu \sqg d\Si_\mu=\int_{x^0} T^{00}\tf   d\tV~~~{\rm where} ~~~ d\tV=\sq{\tga}d^3x.
\lb{325}
\ee
Therefore with $T^{00}\tf=\1k G^{00}\tf$ and (\ref{238}), the gravitational energy $\td E_G$ is now:
\be
\td E_G=Mc^2-\2k\int_{x^0} \lll   \tRR+K^2-K^{kl}K_{kl}  \rrr d\tV.
\lb{326}
\ee
We may again eliminate  $Mc^2$   and obtain the energy density\footnote{Nat\'ario \cite{Na} wrote (\ref{227}), using (\ref{213}), in the form $\na_k\EE^k=\EE^2+\ha f\BB^2-f^{-1}\ka(T_{00}-\ha g_{00}T)$ and called $\EE^2+\ha f\BB^2$ the field energy density because of its analogy with electromagnetism. This formal definition is unrelated to our $\ep_G$ except in   conformastatic spacetimes with metrics   $ds^2=f(dx^0)^2-f^{-1}d\vec r^{~2}$, see below, where the two ``densities" are indeed equal to $\EE^2$.} in terms of a non local field.  Thus, by analogy with (\ref{33}) we set
\be
\tga^*_{kl}= {\rm e}^{2\td\chi}\tga_{kl}~~~{\rm and} ~~~\td\FF_k=-\di_k\td\chi,
\label{327}
\ee
and demand that the   scalar curvature of the conformal metric $\tRR^*=0$. This defines $\td\chi$ by an equation similar to (\ref{36}) or (\ref{37}):
\be
 \tRR= -4\na_k{\td\FF}^k+2{\td\FF}^2~~~{\rm or}~~~\na^2{\rm e}^{\td\chi/2}-\fr{1}{8}\tRR{\rm e}^{\td\chi/2}=0.
 \lb{328}
\ee
The   asymptotic condition on $\td\chi$ is the same as (\ref{38}). If $K=0$ and the dominant energy condition (\ref{18}) is satisfied, then $T^{00}>0$ and thus $\tRR$ which is proportional to $T^{00}+K^{kl}K_{kl}$ is positive. It then follows, see below (\ref{38}), that a solution for $\td\chi$ {\it always exists},  whereas we have not proved this for static observers and, instead of (\ref{326}), we obtain,  
\be
\td E_G=- \1k\int_{x^0}   ( \td \FF^2 -\ha  K^{kl}K_{kl} ) d\tV = \int_{x^0} \td\ep_G d\tV~~~,~~~(K=0).
\lb{329}
\ee 
Note that the $KK$ term has the opposite sign to the $\td\FF^2$ term.  
The quantity $\td\ep_G$ is the gravitational energy density in $\tga$-space, i.e., on $x^0=$ const. For practical calculations of the total gravitational energy, we may use a   procedure  similar to that used in (\ref{313})-(\ref{318}), and obtain $\td E_G$ in terms of local fields:
\be
\td E_G= - \2k\int_{x^0} (\tLL-  K^{kl}K_{kl}) d\tV~~~{\rm in~which}~~~\tLL=\LL(\ga\Ra\tga)~~~,~~~(K=0),
\lb{330}
\ee
and  also a formula similar to (\ref{323}):
\be
\td E_G=- \fr{2}{\ka}\int_{x^0}( \tEE^2-K^{kl}K_{kl}) d\tV-\int (  T^k_k +W^kT^0_k) d\tV~~~,~~~(K=0).
\ee  

\sect{Is gravitational energy negative?}

In Einstein's theory, gravity  may be repulsive.\footnote{This follows from Raychaudhury's equation \cite{Ra}. See Ellis and van Elst \cite{EV} for a covariant presentation (on the web). Repulsive gravity is discussed in various places see, for instance,   \cite{KL}.} In  Newton's theory, gravity is always attractive and gravitational energy is negative. In relativistic gravity we   know, for instance see \cite{Ka2}, that  even in the linearized theory of gravity, $E_G$ is    negative   when the source of gravity is a  perfect fluid but we do not know if it is true otherwise. Nevertheless, there are   important instances in which $E_G<0$ and some of these are considered here.
\vs
\nnn{\it (i) Static spacetimes and static observers} 
\vs
In static \ssss $\AAA=0$ and, see (\ref{222}), $\RR=2\ka T^0_0$. The energy condition (\ref{18}) is here $T_{00}=fT^0_0>0$ and implies $\RR>0$; in this case, as already shown by Cantor \cite{Ca}, a solution of (\ref{37}) exists and
\be
E_G=\td E_G=- \1k\int \LL  dV=-\1k\int \FF^2 dV<0.
\label{41}
\ee
 In spherical symmetry, the  3-space is conformally flat and the metric is of the form
 \be
 ds^2=f(dx^0)^2-bd\vr^{~2}.
 \lb{41b}
 \ee
$f$ and $b$ are functions of $r$.  The case has been studied in some detail in \cite{Ka2} where the connection with MTW's formula for gravitational energy has been given explicitly.
 Using $\LL$ to calculate $E_G$ we readily find that
 \be
 E_G=-\1k\int[(b^{-1/2})']^2dV.
 \lb{41c}
 \ee
 Since outside matter,  
  \be
  b = \lll1+\fr{m}{2r}\rrr^4,
 \lb{42}
 \ee
 any spherically symmetric stellar model of isotropic radius $r_M$ has a total energy  
 \be
 E_{star}=Mc^2-E_G(r\ge r_M)=Mc^2+\fr{GM^2}{2 r_M}=Mc^2\lll 1+\fr{m}{2r_M} \rrr.
 \lb{43}
 \ee
 In the limiting case of a Schwarzschild black hole,   $r_M=m/2$ at the horizon and  the energy of the hole itself is $2Mc^2$. This is    the ``quasi-local" energy  attributed to the black hole by Brown and York \cite{BY}.
 \vs

\nnn{\it (ii) Conformastationary spacetimes for static observers}
\vs
In conformastationary spacetimes \cite{ES},  $\ga_{kl}$ is itself conformally flat but with no special symmetry. If we replace $f$ by $ {\rm e}^{2\psi}$, see (\ref{223}), the metric takes the following form:
  \be
 ds^2=  {\rm e}^{2\psi} \lll dx^0-\AAA_kdx^k\rrr^2 - {\rm e}^{-2\chi}\sum\limits_k (dx^k)^2.
\label{44}
\ee
$\RR$   itself has the form of (\ref{36}) and $E_G$ is given by (\ref{312}) in which $\FF^2$ is now purely local. 

The special case in which $\chi=\psi $ and thus $\FF_k=\EE_k$ has been the
object of much scrutiny in empty spacetimes\footnote{ See \cite{ES} for a review of
the subject and references to original works. Solutions belong to  a class of Einstein-Maxwell  equations. With  $\na_k\AAA^k=0$, the metric is in harmonic coordinates, see \cite{BK}.} and in spacetimes with sources \cite{KBL99}. The gravitational energy for  these metrics  has been studied in some detail in \cite{Ka2}; the following additional remarks are of some interest.\footnote{Beware of the definition of $\BB$ in \cite{Ka2} which is   slightly different.} If $\FF_k=\EE_k$,  (\ref{312}) can be written in the following form which is slightly more complicated but, as can be seen, it shows that $E_G$ is manifestly negative:
\be
E_G=-\1k\int f [   \fr{1}{4}     (\BB+  \AAA\times \EE)^2+   \fr{1}{4}(\AAA\cdot\EE)^2  +f^{-1}(1- \fr{1}{4}f\AAA^2)\EE^2 ] dV<0.
\lb{45}
\ee
 It is manifestly negative because, see (\ref{27}),  
\be
g^{00}=f^{-1}(1-f\AAA^2)=\fr{det(g_{kl})}{g}>0.
\lb{46}
\ee
 Expression (\ref{45}) gives a purely local gravitational energy density when expressed in spatial conformally flat coordinates.

It is interesting to rewrite the last expression in a form which is   similar to the energy density of the electromagnetic field. With
\be
\AAA^*=\ha f^{1/2}\AAA~~~,~~~\BB^*=\na\ti\AAA^*~~~{\rm and}~~~\EE^*=-\ha\na f^{1/2},
\lb{47}
\ee
(\ref{45}) has this form
\be
E_G=-\1k\int\{ {\BB^*}^2+4f^{-1}[ (\AAA^*\cd\EE^*)^2 + (1-{\AAA^*}^2){\EE^*}^2 ]\} dV<0.
\lb{48}
\ee

There is a striking resemblence between  $E_G$ and the energy of the electromagnetic field $E_{EM}$. The electromagnetic energy momentum tensor is, in standard notation:
\be
T_{EM}^{\mu\nu}=\fr{1}{4\pi}( F^{\mu\rho}F_\rho^{~\nu}+\fr{1}{4}g^{\mu\nu}F^{\rho\si}F_{\rho\si}) ~~~{\rm where}~~~F_{\mu\nu}=\di_\mu A_\nu-\di_\nu A_\mu,
\lb{49}
\ee 
and so 
\be
E_{EM}=\int \sq{-g}(T^\mu_\nu)_{EM}w^\nu d\Si_\mu=\int (T^0_0)_{EM}dV.
\lb{410}
\ee 
The electric field components are   
 $E_k= F_{k0}=\di_kA_0$ while those of the magnetic field are  
$B^k= \eta^{kmn}\di_mA_n$. In those terms $E_{EM}$ takes the following form in our coordinates (\ref{23}):
\be
E_{EM}=\fr{1}{8\pi}\int  \{ B^2 + 4f^{-1} [ (\AAA^*\cdot E)^2+  (\fr{1}{4}-{\AAA^*}^2)E^2 )]\}dV>0,
\lb{411}
\ee
where $B^2= \ga_{kl}B^kB^l$ etc...\,\,. Notice that in (\ref{411})  $\AAA^*$ is the gravitational  ``vector potential" defined in (\ref{47}), not the electromagnetic vector potential $A$.
\vs
\nnn{\it (iii) Axially symmetric \ssss  and zero angular momentum observers}
\vs
In axial symmetry, the metric (\ref{232}) takes this special form:
\be
ds^2=\tf (dx^0)^2-a\sum\limits_{K=1,2}(dx^K)^2-b(d\vf -\om dx^0)^2;
 \lb{412}
\ee 
$\tf, a, b$ and $ \om =d\vf/dx^0$,   the   angular velocity of the zero angular momentum observers as measured at infinity, are all functions of $x^K~(K, L, M,\cdots=1,2)$. Under  these conditions, the second fundamental form  components are:
\be
K_{LM}=K_{33}=0~~~{\rm and}~~~K_{L3}=\ha\tf^{-1/2} b\di_L\om.
\lb{413}
\ee
 Consequently,  the gauge condition $K=0$ holds and   $\td\FF$ exists;  using    (\ref{329}),  we find\footnote{Notice that $\na\om\cdot\na\om=a^{-1}\sum\limits_{K=1,2}(\di_K\om)^2$.} that
 \be
 \td E_G=-\1k\int(\td\FF^2  - \fr{1}{4}  \tf ^{-1}b  \na\om\cdot\na\om) d\tV.
 \lb{414}
 \ee
  A particular case which gives a simple expression for $\td E_G$ is when the $\tga$-metric is  conformally flat, i.e. $b/R^2=a$, and 
  \be
  ds^2=\tf (dx^0)^2-a\LLL dR^2+dz^2+R^2(d\vf -\om dx^0)^2\RRR. 
\lb{415}
  \ee

 This metric may be written in the following form:
  \ba
 ds^2&=&f_A(dx^0-\AAA_A Rd\vf)^2-a\LLL dR^2+dz^2+(1+a^{-1}f_A\AAA_A^2)R^2d\vf^2\RRR,\lb{416}\\
 f_A&=&\tf-aR^2\om^2 ~~{\rm and}~~\AAA_A=-f_A^{-1}aR\om,
  \lb{417}
 \ea
which makes it clear   that it is not  a conformastationary metric.
  To calculate the total \gre we better use (\ref{330}) with $\td\LL$ rather than $\td\FF$ and a `background' metric
 \be
 d\Bar s^2=(dx^0)^2- dR^2-dz^2-R^2d\vf ^2.           
 \lb{418}
 \ee
 The \gre is 
 \be
 \td E_G= - \fr{1}{2\ka}\int( \na\log a\cdot\na\log a -\ha \tf^{-1}b\na\om\cdot\na\om )d\tV. 
 \lb{419}
 \ee
The ZAMOs see less kinetic energy since they are dragged in the direction in which the   source rotates and, therefore, one expects gravitational energy to be less negative. It is surely the case for slowly rotating sources,  and presumably the case for fast rotating ones. 

One may use (\ref{330}) to calculate the energy of an axially symmetric distribution of matter and, in the limit, of a Kerr black hole.  An expression for quasilocal energy of a Kerr black hole, based on  paper  \cite{BY},  has been given by Martinez \cite{Ma}. There exists a great number of quasilocal expressions for energy which give different expressions for the energy of  Kerr black holes \cite{Be}.  
\vs 
 \begin{appendix} {\Large {\bf Appendix: Gravitational energy and conservation laws}}
 \vs
A   different derivation of (\ref{318}), closely related to classical conservation laws in general relativity, was given in \cite{Ka2} and the result appears at first strikingly different from what is given here. The derivation of the original formula is very straightforward if one deals with stationary spacetimes.  

We start from the old Freud\footnote{A contemporary accessible version on the web can be found in \cite{PK}.}\cite{Fr} complex $\hat t^{\mu\nu}_{~~\la}$ antisymmetric in the two upper indices. In the notations\footnote{The notations are those of the present paper with some additions: multiplication by $\sqg$ of a tensor like $g^{\mu\nu}$ is represented by a hat like $\hat g^{\mu\nu}=\sqg g^{\mu\nu}$ and, for a change, $\check g^{\mu\nu}=g^{\mu\nu}/\sqg$ while $\Bar D$ stand for 4-covariant derivatives in the $\Bar g$-background spacetime.} of \cite{Ka2} where a background metric is used we have
\ba
\hat t^{\mu\nu}_{~~~\la}= \1k \check g_{\la\rho}\Bar D_\si (\hat g^{\rho[\mu}\hat g^{\nu]\si}). 
\nn
\ea
A remarkable property of the tensor is this \cite{PK}: multiply by any vector field, say, $h^\la$; the divergence of the resulting anti-symmetric tensor density,   a conserved or divergenceless vector field density, is  of the following form:
\ba
\di_\nu(\hat t^{\mu\nu}_{~~~\la}h^\la)= (\hat T^\mu_\nu  + \hat t^\mu_{~\nu}) h^\nu+ \hat t^{\mu\nu}_{~~~\la}\Bar D_\nu h^\la; 
\nn
\ea
$T^\mu_\nu$ is the matter tensor and $t^\mu_{~\nu}$ is Einstein's energy ``complex" in Rosen's \cite{Ro} covariantized form. If $h^\la$ is replaced by the   field of, for example, observers  at rest [see  (\ref{21})], the flux of $\hat t^{\mu\nu}_{~~~\la}w^\la$ through a sphere at infinity, as shown by Freud, is
\ba
\int \hat t^{\mu\nu}_{~~~\la}w^\la dS_{\mu\nu}=\int_{r\ra\in}\sq{\ga} t^{0l}_{~~0}dS_l=Mc^2.
\nn
\ea
Since the volume integral of $\hat T^\mu_\nu w^\nu$ is equal to the mechanical energy $E_M$, it follows that gravitational energy is now given by the following expression:
\ba
E_G=\int  (\hat t^\mu_{~\nu} w^\nu+ \hat t^{\mu\nu}_{~~~\la}\Bar D_\nu w^\la)d\Si_\nu =\int \lll  t^0_{~0}+t^{0l}_{~~0}\EE_l \rrr dV.
\nn
\ea
It takes some more calculations  to show that, if the metric is of the form (\ref{23}), this integral is indeed   (\ref{318}).  
\end{appendix}
\vs
 \Large{\bf Acknowledgements}
 \vs
 \normalsize 
 \setlength{\baselineskip}{20pt plus2pt}
 We acknowledge the support of the grant GA\v CR 202/06/0041 of the Czech Republic, the Royal Society Grant (JB, DLB),  and the hospitality of the Institute of Astronomy, Cambridge (JB, JK).
  
 JK is grateful to Nathalie Deruelle for interesting and useful remarks.
 
  \end{document}